\documentclass[12pt]{iopart}

\usepackage{bm}
\usepackage{epsfig}
\usepackage{citesort}
\usepackage{graphicx}
\eqnobysec

\newcommand{\lwig}{\mbox{\;\raisebox{.3ex}
    {$<$}$\!\!\!\!\!$\raisebox{-.9ex}{$\sim$}\;}}

\newcommand{\lambdabar}%
{{\hbox{$\lambda$\kern-1.ex\raise+0.45ex\hbox{--}}}}

\begin{document}


\begin{flushright}
{\large \tt MPP-2010-36\\
TTK-10-28}
\end{flushright}

\title{Neutrino and axion hot dark matter bounds after WMAP-7}

\author{Steen~Hannestad$^1$, Alessandro Mirizzi$^2$,
Georg~G.~Raffelt$^3$ and
Yvonne~Y.~Y.~Wong$^4$}

\address{
 $^1$~Department of Physics and Astronomy\\
 University of Aarhus, DK-8000 Aarhus C, Denmark\\
$^2$~II.~Institut f\"ur Theoretische Physik, Universit\"at Hamburg \\
Luruper Chaussee 149, D-22761 Hamburg, Germany\\
 $^3$~Max-Planck-Institut f\"ur Physik (Werner-Heisenberg-Institut)\\
 F\"ohringer Ring 6, D-80805 M\"unchen, Germany\\
$^4$~Institut f\"ur Theoretische Teilchenphysik und Kosmologie\\
 RWTH Aachen, D-52056 Aachen, Germany}

\ead{\mailto{sth@phys.au.dk}, \mailto{alessandro.mirizzi@desy.de},
     \mailto{raffelt@mppmu.mpg.de},
     \mailto{yvonne.wong@physik.rwth-aachen.de}}

\date{\today}

\begin{abstract}
We update cosmological hot dark matter constraints on neutrinos and
hadronic axions. Our most restrictive limits use 7-year data from
the Wilkinson Microwave Anisotropy Probe for the cosmic microwave
background anisotropies, the halo power spectrum (HPS) from the 7th
data release of the Sloan Digital Sky Survey, and the Hubble
constant from Hubble Space Telescope observations. We find 95\% CL
upper limits of $\sum m_\nu<0.44$~eV (no axions), $m_a<0.91$~eV
(assuming $\sum m_\nu=0$), and $\sum m_\nu<0.41$~eV and
$m_a<0.72$~eV for two hot dark matter components after marginalising
over the respective other mass. CMB data alone yield $\sum
m_\nu<1.19$~eV (no axions), while for axions the HPS is crucial for
deriving $m_a$ constraints. This difference can be traced to the
fact that for a given hot dark matter fraction axions are much more
massive than neutrinos.
\end{abstract}

\maketitle

\section{Introduction}                        \label{sec:introduction}

Cosmological large-scale structure data allow for precise estimates
for the parameters of minimal or extended cosmological models. These
results have ramifications far beyond cosmology itself, notably in
the area of neutrino physics. The well-known hot dark matter
constraints provide neutrino mass limits that directly impact
neutrino mass searches in single~\cite{Drexlin:2008zz} and
double~\cite{Schonert:2010zz} beta decay experiments. In a series of
papers by our
collaboration~\cite{Hannestad:2003ye,Hannestad:2005df,Hannestad:2007dd,Hannestad:2008js}
and another group~\cite{Melchiorri:2007cd} this approach was
extended to hadronic axions where the resulting mass limits are
complementary to solar axion searches by the CAST
experiment~\cite{Zioutas:2004hi,Andriamonje:2007ew,Arik:2008mq} and
the Tokyo axion
helioscope~\cite{Moriyama:1998kd,Inoue:2002qy,Inoue:2008zp}. In the
present work, we use the 7-year data release from the Wilkinson
Microwave Anisotropy Probe (WMAP) as an opportunity to update these
results, and also modify along the way several other input assumptions
as detailed in the main text below.

\begin{figure}[t]
\hspace{25mm}
\includegraphics[width=12.5cm]{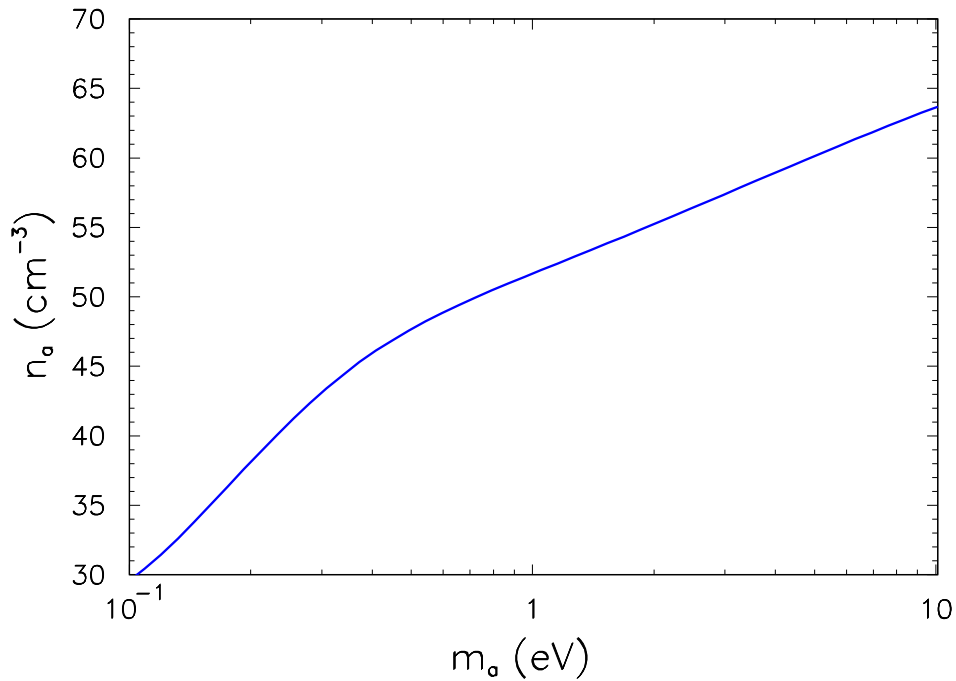}
\caption{Axion number density $n_a$ as a function of the axion mass
$m_a$ assuming the hadronic thermalisation processes~(\ref{eq:axionthermalisation})
based on the calculations of reference~\cite{Hannestad:2005df}.\label{fig:na}}
\end{figure}

Within standard cosmological assumptions, the neutrino plus
antineutrino number density today, summed over all flavours, is
$n_\nu\sim336~{\rm cm}^{-3}$. Currently available cosmological data
are not yet sensitive enough to resolve the small mass differences
measured in oscillation experiments, so all neutrinos are treated as
having the same mass $m_\nu$, traditionally expressed by the
parameter $\sum m_\nu=3 m_\nu$. For axions, on the other hand, the
freeze-out temperature and therefore the number density $n_a$
depends on the axion's interaction rate with pions and nucleons via 
\begin{eqnarray}
\label{eq:axionthermalisation}
a+\pi &\leftrightarrow& \pi +\pi, \nonumber \\
a+N & \leftrightarrow& N + \pi,
\end{eqnarray}
where the coupling strength is, in turn, proportional to the axion mass $m_a$
\cite{Berezhiani:1992rk,Chang:1993gm,Hannestad:2005df}. 
Figure~\ref{fig:na} shows the relation between $m_a$ and $n_a$
computed for the thermalisation processes~(\ref{eq:axionthermalisation})
based on the original calculations of reference~\cite{Hannestad:2005df}.
For small
$m_a$, the number density is also small, so assuming $m_a=0$ implies
$n_a=0$ which brings us back to standard cosmology. Near the hot
dark matter limit of $m_a\sim 1$~eV one finds a present-day number
density of $n_a\sim 50~{\rm cm}^{-3}$. Therefore, in the relevant
mass range, neutrinos are about 6 times more numerous than axions
and one expects a hot dark matter limit on $m_a$ that is roughly
twice that on $\sum m_\nu$, in agreement with what we find when we
use our full range of input data sets.

In detail, however, the situation is more subtle. Conventional
wisdom says that hot dark matter constraints arise primarily from
the shape of the measured matter power spectrum, since hot dark
matter free-streaming suppresses the growth of matter perturbations
and hence the clustering power on small scales. However, recent
cosmological data have become so precise that one finds a useful
limit on $\sum m_\nu$ of order 1~eV already from the cosmic
microwave background (CMB) anisotropies alone, notably from the
increased amplitude of the first acoustic peak in the temperature
auto-correlation spectrum due to the early integrated Sachs--Wolfe
(ISW) effect~\cite{Ichikawa:2004zi}. The same is however not true
for axion hot dark matter, since for the same hot dark matter
fraction, the axion is necessarily some 6 times heavier than the
equivalent neutrino.  Thus while neutrinos with masses near the hot
dark matter limit ($m_\nu \lwig 0.3~{\rm eV}$) essentially act like
radiation at CMB decoupling and contribute strongly to the early ISW
effect, the equivalent axion is already nonrelativistic and thus
indistinguishable from cold dark matter as far as the CMB is
concerned.  In the latter case, one needs to use the shape of the
matter power spectrum from smaller-scale data in order to put any
sensible constraint on $m_a$.

In order to derive new hot dark matter limits on neutrinos and
axions and to explain their differences, we begin in
section~\ref{sec:model} with a description of our cosmological model
and in section~\ref{sec:data} of the data sets used. In
section~\ref{sec:results} we use standard Bayesian techniques to derive
credible intervals for $\sum m_\nu$ and $m_a$ separately and for a
two-component case based on different combinations of data sets. We
discuss and summarise our findings in section~\ref{sec:conclusions}.

\section{Cosmological model}                         \label{sec:model}

We consider a cosmological model with vanishing spatial curvature and
adiabatic initial conditions, described by eight free parameters,
\begin{equation}\label{eq:model}
{\bm \theta} = \{\omega_{\rm cdm},\omega_{\rm b},H_0,\tau,
\ln(10^{10}A_{\rm s}),n_{\rm s},
\sum m_\nu,m_a\}.
\end{equation}
Here, $\omega_{\rm cdm}=\Omega_{\rm cdm} h^2$ is the physical cold
dark matter density, $\omega_{\rm b}=\Omega_{\rm b} h^2$ the baryon
density, $H_0=h~100~{\rm km~s^{-1}~Mpc^{-1}}$ the Hubble parameter,
$\tau$ the optical depth to reionisation, $A_{\rm s}$ the amplitude
of the primordial scalar power spectrum, and $n_{\rm s}$ its
spectral index. These six parameters represent the simplest
parameter set necessary for a consistent interpretation of the
currently available data.

In addition, we allow for a nonzero sum of neutrino masses $\sum
m_\nu$ and a nonvanishing axion mass $m_a$. These
extra parameters will be varied one at a time, as well as in
combination. Their ``standard'' values are given in
table~\ref{tab:priors}, along with the priors for all cosmological
fit parameters considered here.

\begin{table}[t]
\caption{Priors and standard values for the cosmological fit
parameters considered in this work.  All priors are uniform (top
hat) in the given intervals.\label{tab:priors}} \hskip25mm
{\footnotesize
\begin{tabular}{lll}
\br
Parameter& Standard & Prior\\
\mr
$\omega_{\rm cdm}$   & ---   & $0.01$--$0.99$ \\
$\omega_{\rm b}$    & ---   & $0.005$--$0.1$ \\
$h$                 & ---   & $0.4$--$1.0$\\
$\tau$              & ---   & $0.01$--$0.8$ \\
$\ln(10^{10}A_{\rm s})$   & ---   & $2.7$--$4.0$ \\
$n_{\rm s}$               & ---   & $0.5$--$1.5$ \\
$\sum m_\nu$ [eV]       & 0 & $0$--$10$ \\
$m_a$ [eV]      & 0     & $0$--$10$ \\
\br
\end{tabular}
}
\end{table}

\section{Data}                                        \label{sec:data}

\subsection{Cosmic microwave background (CMB)}

We use a compilation of measurements of the CMB temperature and
polarisation anisotropies from WMAP after seven years of
observation~\cite{Larson:2010gs,kom10},
ACBAR~\cite{Reichardt:2008ay}, BICEP~\cite{Chiang:2009xsa}, and
QuAD~\cite{Brown:2009uy}.

\subsection{Halo power spectrum (HPS)}

We use the halo power spectrum constructed from the luminous red
galaxy (LRG) sample of the seventh data release of the Sloan Digital
Sky Survey (SDSS-DR7)~\cite{Reid:2009xm}. The full HPS data set
consists of 45 data points, covering wavenumbers from $k_{\rm min} =
0.02\ h {\rm Mpc}^{-1}$ to $k_{\rm max} = 0.2\ h {\rm Mpc}^{-1}$. We
fit this data set following the procedure of
reference~\cite{Reid:2009xm}, using a properly smeared power
spectrum to model nonlinear mode coupling. The smearing procedure
requires that we supply a smooth, no-wiggle power spectrum, which we
construct using the discrete spectral analysis technique introduced
in reference~\cite{Hamann:2010pw}.

\subsection{Baryon acoustic oscillations (BAO)}

The baryon acoustic oscillation scale has been extracted from
SDSS-DR7~\cite{Percival:2009xn}, which provides an angular diameter
distance measure at $z=0.275$. However, since parameters like $\sum
m_\nu$ and $m_a$ can in principle affect the acoustic scale, care
needs to be taken when evaluating the BAO likelihood. We refer the
reader to reference~\cite{Hamann:2010pw} for a more detailed
discussion of this issue.

\subsection{Hubble parameter from the Hubble Space Telescope (HST)}

We adopt the constraint on the Hubble parameter derived from
observations with the Hubble Space Telescope~\cite{Riess:2009pu}.

\section{Results}                                  \label{sec:results}

\begin{table}[t]
\caption{1D marginal 95\% upper bounds on  $\sum m_\nu$ and $m_a$ for
several different choices of data sets and models.
\label{tab:results}} \hskip25mm {\footnotesize
\begin{tabular}{llcc}
\br
Model & Data set  & $\sum m_\nu$ [eV] & $m_a$ [eV]\\
\mr
Fixed $m_a=0$ &CMB only &  1.19 & --- \\
&CMB+BAO & 0.85 & --- \\
&CMB+HST & 0.58 & --- \\
&CMB+HPS & 0.61 & --- \\
&CMB+HPS+HST & 0.44 & --- \\
\mr
Fixed $\sum m_\nu=0$ & CMB only & --- & No constraint \\
&CMB+BAO & --- & No constraint \\
&CMB+HST  & ---& No constraint \\
&CMB+HPS &  --- & 1.07 \\
& CMB+HPS+HST & --- & 0.91\\
\mr
Free $\sum m_\nu$ and  $m_a$  &CMB+HPS  & 0.58 & 0.82    \\
&CMB+HPS+HST   & 0.41 & 0.72    \\
\br
\end{tabular}
}
\end{table}

\begin{figure}[t]
\hspace{25mm}
\includegraphics[height=12.cm,angle=-90]{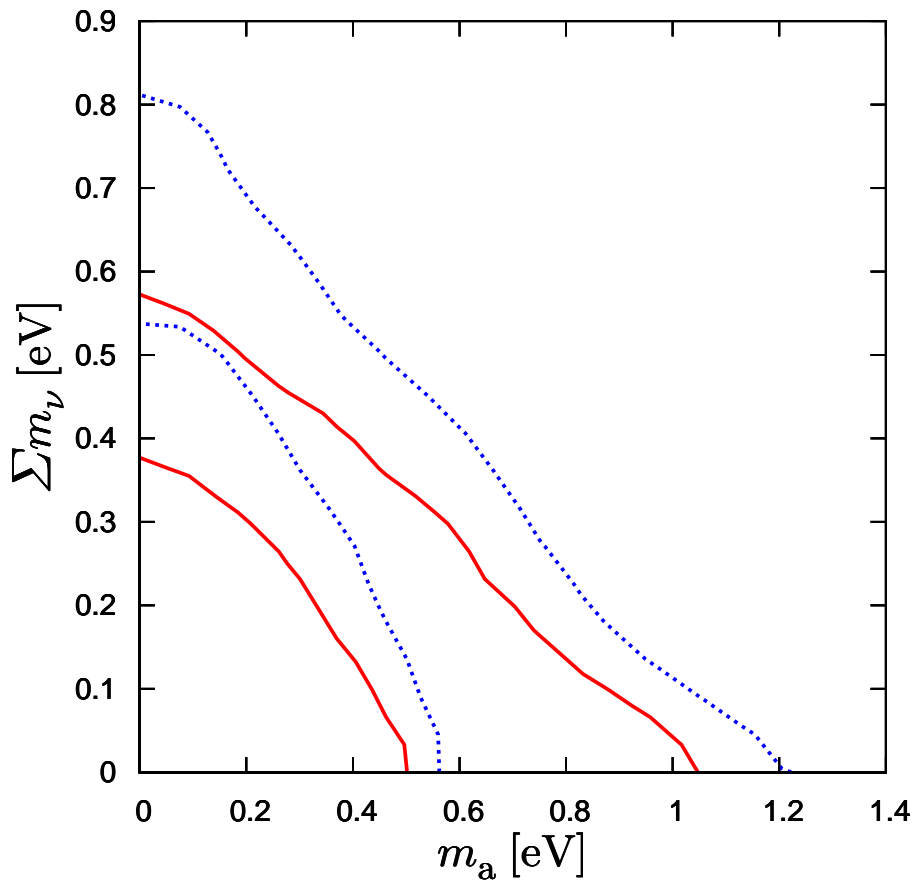}
\caption{2D marginal 68\% and 95\% contours in the $\sum m_\nu$--$m_a$ plane.
The blue lines correspond to  our  results using CMB+HPS, and the
red lines using CMB+HPS+HST. \label{fig:contours}}
\end{figure}

We use standard Bayesian inference techniques and explore the model
parameter space with Monte Carlo Markov Chains (MCMC) generated
using the publicly available {\sc CosmoMC}
package~\cite{Lewis:2002ah}.  Our results are summarised in
table~\ref{tab:results} and figure~\ref{fig:contours}.

In agreement with several recent papers, the CMB alone provides a fairly robust
limit on $\sum m_\nu$.   With $m_a$ held fixed at zero,
we find $\sum m_\nu < 1.2$ eV at 95\% C.L.\ using WMAP and other CMB data,
while Komatsu {\it et~al.} find $\sum m_\nu < 1.3$ eV at 95\% C.L.\ from WMAP alone~\cite{kom10}.

It is noteworthy that for the opposite case when $m_\nu$ is fixed at
zero, CMB data provide no constraint on $m_a$. At first sight this might
seem contradictory, but there is a simple explanation. The
suppression of small scale power is essentially controlled by the
fraction of hot to cold dark matter, $f_{\rm hdm} = \omega_{\rm
hdm}/\omega_{\rm cdm}$, for both axions and neutrinos. This can be
seen in figure~\ref{fig:pk} where the red/solid and green/dash lines have the
same asymptotic behaviour at large $k$.  However, the number density of axions is
six times smaller than that of neutrinos.  Therefore in order to give
the same contribution to $\omega_{\rm hdm}$, the mass of the axion must be
correspondingly larger than the neutrino mass.

\begin{figure}[t]
\hspace{25mm}
\includegraphics[height=12.5cm,angle=-90]{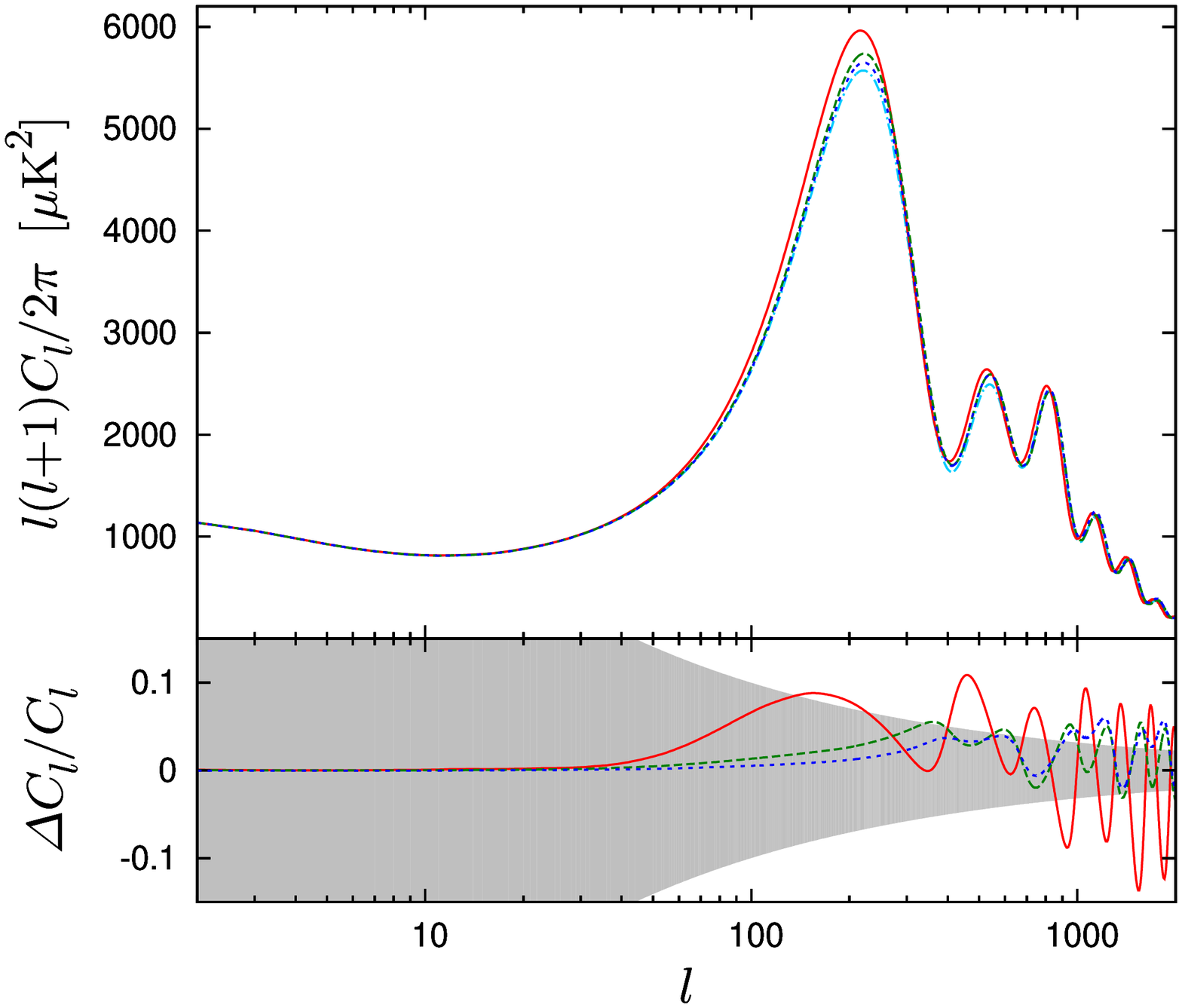}
\caption{CMB temperature anisotropy spectra
for various mixtures of cold and hot dark matter, where
$\omega_{\rm cdm}+\omega_\nu+\omega_a=0.112$ is held constant.
Bottom: Differences to standard $\Lambda$CDM; the
shaded/grey area indicates cosmic variance. Light blue/dot-dash
line: standard $\Lambda$CDM ($\omega_{\rm cdm}=0.112$, $\omega_a=0$,
$\omega_\nu=0$).
  Red/solid line: $\Lambda$CDM+$\nu$ with $\sum m_\nu = 1.2~{\rm eV}$
 ($\omega_{\rm cdm}=0.099$, $\omega_\nu=0.013$). Green/dash line:
$\Lambda$CDM+$a$ with $m_a = 2.4~{\rm
eV}$ ($\omega_{\rm cdm}=0.099$, $\omega_a=0.013$). Dark blue/dotted
line: Extreme axion case with $m_a = 10~{\rm eV}$
($\omega_{\rm cdm}=0.0498$, $\omega_a=0.0622$).
\label{fig:cls}}
\end{figure}

\begin{figure}[t]
\hspace{25mm}
\includegraphics[height=12.5cm,angle=-90]{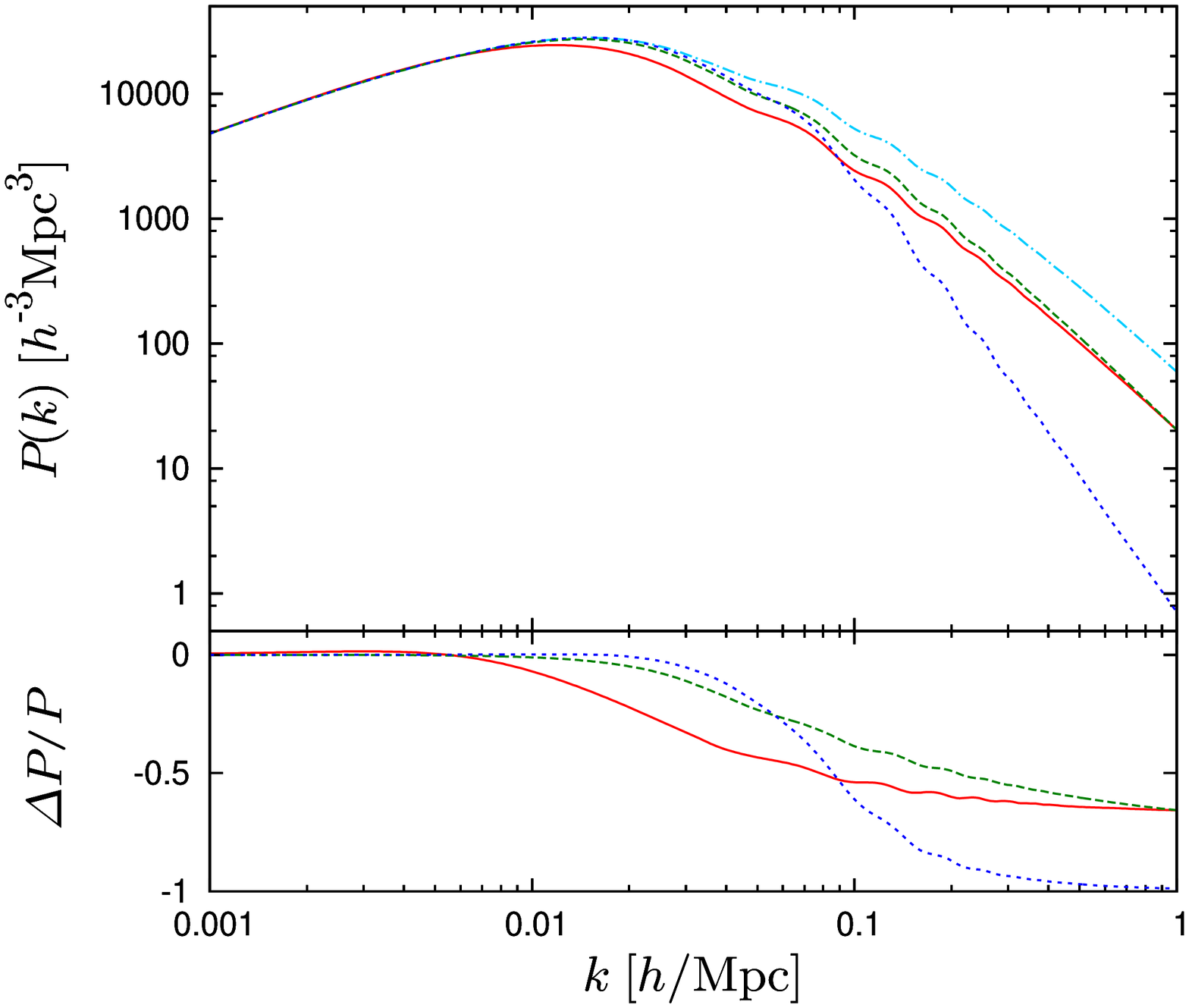}
\caption{Same as figure~\ref{fig:cls},
but for the matter power spectrum.\label{fig:pk}}
\end{figure}

In two examples shown in figures~\ref{fig:cls} and~\ref{fig:pk},
both the neutrino and the axion hot dark matter models have
$\omega_{\rm hdm} = 0.013$, but the corresponding particle masses
are $\sum m_\nu = 1.2$~eV (or $m_\nu = 0.4$~eV) and $m_a \simeq
2.4$~eV respectively. The temperature at recombination is
approximately 0.2~eV.  A neutrino with $m_\nu = 0.4$~eV would still
be semi-relativistic at this time and contribute significantly to
the early ISW effect. At $m_a \simeq 2.4$~eV, however, the axion is
nonrelativistic at recombination and behaves more like cold dark
matter. This disparity between the two hot dark matter candidates
can be clearly seen in figure~\ref{fig:cls}, where the CMB
temperature anisotropy spectrum for the axion model is almost
indistinguishable from standard $\Lambda$CDM, while its neutrino
counterpart shows a pronounced early ISW effect.

The same effect is also manifest in the matter power spectrum shown
in figure~\ref{fig:pk}, where the epoch of matter--radiation
equality---marked by the turning point of the spectrum---is
different for axions and neutrinos. In the neutrino case,
matter--radiation equality happens later, leading to a turning point
at a much smaller value of $k$  compared with its $\Lambda$CDM
counterpart.  In the axion case, the corresponding matter power
spectrum traces closely that of the $\Lambda$CDM model up to $k \sim
0.02 \ h \ {\rm Mpc}^{-1}$.  Beyond this point, axion
free--streaming sets in and suppresses the power on small scales
relative $\Lambda$CDM.  The amount of suppression in $P(k)$ at large
$k$ values is similar in both the neutrino and the axion case as
expected, since this is governed primarily by~$f_{\rm hdm}$.

Figures~\ref{fig:cls} and \ref{fig:pk} also show an extreme example
of power spectra for $m_a=10~{\rm eV}$, corresponding to
$\omega_{a}=0.0622$, actually exceeding $\omega_{\rm cdm}=0.0498$.
Axions with such a large mass act essentially as cold dark matter as
far as CMB anisotropies are concerned.

In practice this means that when only CMB data are used, it is
difficult to distinguish a several-eV axion model from the standard
$\Lambda$CDM model, and accordingly WMAP provides no useful upper
bound on $m_a$. This is true even when additional priors are
imposed, either in the form of the BAO scale, or the HST measurement
of the Hubble parameter. Only when low-redshift small-scale data
containing {\it shape} information such as the HPS are added does an
upper bound on $m_a$ emerge. This is an example of the point
discussed in reference~\cite{Hamann:2010pw}, i.e., that it is
crucial to extract the full matter power spectrum from a large scale
structure survey because it contains important information not
stored in the geometric BAO scale.

From our most restrictive data sets and allowing both the neutrino
and the axion masses to vary simultaneously, we find for the axion
mass $m_a < 0.72$ eV (95\% C.L.), after marginalising over all other
model parameters including $\sum m_\nu$. The corresponding 2D
marginalised 68\% and 95\% contours in the $\sum m_\nu$--$m_a$ plane
are shown in figure~\ref{fig:contours}.  This new limit on $m_a$ is
somewhat tighter than our previously published bound of $m_a <
1.02$~eV (95\% C.L.) using the 5-year WMAP
data~\cite{Hannestad:2008js}. Fixing $\sum m_\nu$ to zero, the limit
on $m_a$ becomes less restrictive ($m_a < 0.91$~eV at 95\% C.L.),
which would be applicable if future laboratory experiments were to
provide a significant constraint on $m_\nu$.   On the other hand, if
these experiments were to measure a value of $m_\nu$ close to the
current cosmological limit, then less room would be left for axion
hot dark matter and the limit on $m_a$ would tighten
correspondingly.

\section{Conclusions}                          \label{sec:conclusions}

We have provided an updated constraint on axion hot dark matter
using new cosmological data, most notably CMB data from the WMAP
7-year data release and the final SDSS-DR7 LRG data. We have also
pointed out a qualitative difference between neutrino and axion hot
dark matter, in that axion masses in the detectable range are large
enough that axions are nonrelativistic at recombination. This means
they act almost like cold dark matter as far as CMB is concerned and
that current CMB data in itself does not provide a useful limit on
the axion mass, even when priors from HST or BAO are imposed. In
order to properly constrain $m_a$ it is necessary to include
information on the shape of the matter power spectrum from, for
example, the SDSS halo power spectrum.

An interesting question is whether future CMB data will be sensitive
to axion hot dark matter. The Planck mission has an estimated
sensitivity of \hbox{$\sigma(\sum m_\nu) \sim 0.3$--0.5~eV}
\cite{DeBernardis:2009di,Lesgourgues:2006nd}.  But from
figure~\ref{fig:cls} we can already say that even at high multipoles
the effect of axion hot dark matter on the CMB anisotropy spectrum
is quite small---generally smaller than the uncertainty due to
cosmic variance. This leads to the inevitable conclusion that CMB
anisotropy observations will remain poor probes of hot dark matter
in the several eV range.

Cosmological bounds on neutrino and axion masses are nicely
complementary to experimental searches, but cannot replace them. One
caveat concerning our axion results is that we need to assume that
the predicted thermal population was actually produced after the QCD
epoch. In non-standard cosmologies with low reheating temperature
the axion population can be severely suppressed and our bounds would
not apply~\cite{Grin:2007yg}.  In other scenarios a significant
cosmic background of low-mass axions can be produced that remain
relativistic until today and do not form hot dark
matter~\cite{Chun:2000jr}. For neutrinos such caveats are less
relevant because their thermalisation epoch is well probed by
big-bang nucleosynthesis and the presence of radiation with roughly
the right abundance has been confirmed by precision cosmology.

\section*{Acknowledgements}

We acknowledge use of computing resources from the Danish Center for
Scientific Computing (DCSC). In Munich, partial support by the
Deutsche Forschungsgemeinschaft under the grant TR~27 ``Neutrinos and
beyond'' and the Cluster of Excellence ``Origin and Structure of the
Universe'' is acknowledged.

\section*{References}

\end{document}